\documentclass[11pt]{article}
\usepackage{latexsym}
\usepackage{amssymb}
\usepackage{epsf}
\usepackage{amsmath}
\usepackage{slashed}
\usepackage[hypertex]{hyperref}
\usepackage{graphicx}

\newcommand{\psl}{p \hspace{-0.5 em}/}

\begin{document}
\pagestyle{empty}


\hfill TU-873

\vspace{3cm}

\begin{center}

{\bf\LARGE Natural Supersymmetry at the LHC}
\\

\vspace*{1.5cm}
{\large 
Masaki Asano$^a$, Hyung Do Kim$^b$,  
Ryuichiro Kitano$^a$, and Yasuhiro Shimizu$^{a,c}$
} \\
\vspace*{0.5cm}

$^a${\it Department of Physics, Tohoku University, Sendai 980-8578,
 Japan}\\
$^b${\it FPRD and Department of Physics, Seoul National University, 
Seoul, 151-747, Korea}
$^c${\it IIAIR, Tohoku University, Sendai 980-8578,
 Japan}\\
\vspace*{0.5cm}

\end{center}

\vspace*{1.0cm}

\begin{abstract}
{\normalsize
If the minimal supersymmetric standard model is the solution to the
 hierarchy problem, the scalar top quark (stop) and the Higgsino should weigh
 around the electroweak scale such as 200~GeV. A low messenger scale,
 which results in a light gravitino, is also suggested to suppress the
 quantum corrections to the Higgs mass parameters. Therefore the minimal
 model for natural supersymmetry is a system with
 stop/Higgsino/gravitino whereas other superparticles are heavy. We
 study the LHC signatures of the minimal system and discuss the
 discovery potential and methods for the mass measurements.

}
\end{abstract} 

\newpage
\baselineskip=18pt
\setcounter{page}{2}
\pagestyle{plain}
\baselineskip=18pt
\pagestyle{plain}

\setcounter{footnote}{0}

\section{Introduction}\label{sect:Intro}

The LHC experiments have started taking data to aim for the discovery of
new physics. As one of the leading candidates for the new physics,
various models of low energy supersymmetry (SUSY) have been constructed
and their signatures have been studied.
Especially, discovery and mass measurements in ``representative'' models
such as the minimal supergravity and the minimal gauge mediation model
have been extensively studied
\cite{ATLAS-TDR,Abdullin:1998pm,Weiglein:2004hn}.

On the theoretical side, SUSY is motivated as a solution to the
hierarchy problem. The electroweak symmetry breaking via the Higgs
mechanism is naturalized by this space-time symmetry.
However, when we closely look at the ``representative'' models used for
the collider studies, we encounter a naturalness problem in the Higgs
sector.
The experimental lower bound on the Higgs mass requires a heavy scalar
top quark (stop), which results in a delay of the cancellation of
quadratic divergence.

The problem is, however, not model independent even within the context
of the minimal supersymmetric standard model (MSSM). For example, the
problem in the minimal supergravity model is the large logarithm, $\log
M_{\rm Pl} / m_{\tilde t} $ ($m_{\tilde t}$ and $M_{\rm Pl}$ are the
stop mass and the Planck scale, respectively), in the radiative
corrections of the Higgs mass parameter. The logarithm, $\log M_{\rm
mess} / m_{\tilde t}$ ($M_{\rm mess}$ is the messenger scale), can be
small in the minimal gauge mediation model. However the small $A$-term
is the problem in this case; the Higgs boson mass requires a very heavy
stop otherwise the large $A$-term is required.

The conditions for the naturalness have been summarized in
Ref.~\cite{Kitano:2006gv}: (a) light stops, (b) light Higgsinos, (c) a
moderately large value of $\tan \beta$, (d) a large $A$-term and (e) a
small logarithm.

In the discussion of the naturalness of the electroweak symmetry
breaking, the most relevant parameter is the stop mass due to the large
top Yukawa coupling and the color factor. In the MSSM, the stop mass, 
$m_{\tilde t}$ 
(the geometric mean of the two stop masses), should be at most about
600~GeV for the fine tuning to be better than at the level of
10~\%~\cite{Chacko:2005ra, Nomura:2005qg, Kitano:2006gv}. Here, a small
logarithm (the low messenger scale) of $M_{\rm mess} \sim 10$~TeV is
assumed.
Even with such a light stop, the lower bound on the Higgs boson mass
from the LEP-II experiments~\cite{Barate:2003sz} can be satisfied within
the MSSM when the $A$-parameter (the stop-stop-Higgs coupling) is as
large as of order the stop mass~\cite{Kitano:2005wc, Dermisek:2006ey,
Kitano:2006gv, Dermisek:2006qj}. (See \cite{Carena:1995wu, Haber:1996fp}
for the study of the $A$-parameter dependence of the Higgs boson mass in
the MSSM.)
Imposing the absence of significant fine tuning at the level of 10\%,
the naturalness upper bound on $m_{\tilde t}$ (which gets severer from
above for a large $A$-parameter) and a large stop mixing through the large 
$A$-parameter required from the Higgs boson mass constraint imply that the
lighter stop should be lighter than about $400$~GeV for $M_{\rm
mess} \gtrsim 10$~TeV.

Other sfermions and gauginos (gluinos, Winos and Binos) can be as heavy
as a few TeV without affecting the naturalness of the electroweak
symmetry breaking.
On the other hand, as the supersymmetric Higgs mass $\mu$ is added up to
the soft supersymmetry breaking Higgs mass parameters, $\mu$ (the
Higgsino mass) should be of the electroweak scale to be natural.

The superparticle spectrum is important when we discuss signatures at
the Large Hadron Collider (LHC).
If we take the naturalness seriously in supersymmetric models, we
see that there should be significant amount of the stop production at
the LHC.
If other sparticles except for the Higgsinos are heavier than 1 TeV,
superparticle productions will be dominated by lighter stops 
and they all decay into Higgsinos directly.
This is the minimal model of natural SUSY at the LHC, that is simply a
system with a stop and Higgsinos (two neutralinos and a chargino). 
The requirement of the small
logarithm implies a low energy SUSY breaking scenario
(see~\cite{Choi:2004sx, Choi:2005ge, Endo:2005uy, Choi:2005uz} for an
interesting exception), and thus we also include the nearly massless
gravitino, to which the lightest Higgsino can decay. 

In the case where the Higgsino is heavier than the stop, the stop
directly decays into a top quark and a gravitino if open. Those events
look like $t\bar t$ productions. We will not consider such a case in
this paper. See Ref.~\cite{Alwall:2010jc} for a study of such an event 
topology.
In the case where the top quark is not open, the stop
undergoes the three body decay into $b$, $W$ and the gravitino. Such a
case has been studied in Ref.~\cite{Chou:1999zb}.

There are attempts of model-building and studies of collider signatures
with similar motivations. More minimal supersymmetry~\cite{Cohen:1996vb}
has the first two generation sparticles at around 5 to 20 TeV to solve
supersymmetry flavor and CP problems while keeping the naturalness for
the electroweak symmetry breaking. 
In Ref.~\cite{Perelstein:2007nx}, a signature of a large $A$-term at the
LHC has been discussed.
See the partial list of recent related works in~\cite{Giudice:2008uk,
Sundrum:2009gv, Aharony:2010ch, Barbieri:2010pd, Redi:2010yv,
Baer:2010ny}.

In this paper, we study LHC signatures of a model with light stops,
light Higgsinos and a (nearly) massless gravitino.
As we have discussed, it is well-motivated to consider a very light stop
such as $m_{\tilde t} \sim 200$~GeV. With such a light colored particle,
the LHC will be very powerful.
We assume that other superparticles are heavy enough to be ignored or
even absent as particle states by directly coupling to the SUSY breaking
sector (see {\it e.g} \cite{Graesser:2009bu}). 
Such a spectrum is also motivated by the constraints from
CP-violation and flavor physics.
The SUSY events with the largest cross section are therefore a pair
production of the lighter stop. The main decay mode of the stop is into
a bottom ($b$) quark and a chargino. The chargino subsequently decays
into a neutralino and soft jets or leptons which we can ignore due to a
small mass splitting between the charged and the neutral Higgsinos. The
neutralino then decays into a $Z$ boson or a Higgs boson and a
gravitino.

The final state is $2b + 2Z$, $2b + hZ$, or $2b + 2h$ with missing
momentum. We analyze these events and consider methods to find the stop.
We also discuss the measurement of the stop and the Higgsino masses as
well as the discovery of the Higgs boson through this process.

\section{Naturalness upper bound on superparticle masses}

We start with the review of the naturalness in electroweak symmetry
breaking in the MSSM (see, e.g., \cite{Kitano:2006gv} for a more
detailed discussion).
Generally in electroweak symmetry breaking via the Higgs mechanism,
there is a relation between the Higgs boson mass $(m_h)$ and the
quadratic term in the potential (the negative mass squared), $m^2$:
\begin{eqnarray}
 {m_h^2 \over 2 } = - m^2.
\end{eqnarray}
In the MSSM, $m_h$ can be as large as 130~GeV~\cite{Okada:1990vk,
Ellis:1990nz, Haber:1990aw}. For a moderately large value of $\tan
\beta$, electroweak symmetry breaking is mainly due to the vacuum
expectation value of the up-type Higgs field, $H_u$. In this case, the
$m^2$ parameter has two sources:
\begin{eqnarray}
 m^2 = \mu^2 + m_{H_u}^2.
\end{eqnarray}
The first and the second terms are positive and negative, respectively.
A fine tuning is required if there are contribution to $|m^2|$ which are
much larger than $m_h^2 / 2$ $\lesssim (130~{\rm GeV})^2/2$.

An obvious conclusion from above is that the $\mu$ parameter (the
Higgsino mass) cannot be very large. If we measure the fine tuning by
$\Delta^{-1} = m_h^2 / 2 \mu^2$ and require $\Delta^{-1} > 10\%$ for
example, we obtain
\begin{eqnarray}
 |\mu| \lesssim 290~{\rm GeV}.
\end{eqnarray}

Another important contribution is from the stop-top loop diagrams to the
$m_{H_u}^2$ parameter. There are two kinds of interactions in the
diagrams; one with the top Yukawa interaction ($- y_t \overline{q}_3 u_3
H_u$) and another through the three-point stop-stop-Higgs interaction (
$- A_t y_t \tilde q_3 \tilde u_3H_u$).  The three-point interaction also
provides an important contribution to the Higgs boson mass, $m_h$.
By assuming a small logarithm ($M_{\rm mess} \sim 10$~TeV), we obtain
upper and lower bounds on the stop mass parameter ($m_{\tilde t} \equiv
(m_{\tilde q_3} m_{\tilde u_3})^{1/2}$) from the naturalness and the
Higgs boson mass bound, respectively.
Requiring $\Delta^{-1} > 10\%$ and $m_h >
114.4$~GeV~\cite{Barate:2003sz}, where $\Delta^{-1} \equiv m_h^2 / 2
m_{H_u}^2|_{\rm rad}$ with $m_{H_u}^2|_{\rm rad}$ the stop-loop
contribution, the bounds are
\begin{eqnarray}
500~{\rm GeV} \lesssim m_{\tilde t} \lesssim 500~{\rm GeV},
\end{eqnarray}
for $|A_t| \sim m_{\tilde t}$ and 
\begin{eqnarray}
250~{\rm GeV} \lesssim m_{\tilde t} \lesssim 360~{\rm GeV},
\end{eqnarray}
for $|A_t| \sim 2 m_{\tilde t}$. There is no allowed region for $|A_t|
\lesssim m_{\tilde t}$ or $|A_t| \gtrsim 2.5 m_{\tilde t}$. For $A_t = 0$,
we obtain $\Delta^{-1} \lesssim 2\% $\footnote{The bounds and values
are sensitive to the top quark mass. We have used $m_t =
173$~GeV~\cite{top:1900yx}.}.
The maximum value of $\Delta^{-1}$ is about 20\% which can be achieved
when $|A_t| \sim 2 m_{\tilde t}$.
The naturalness upper bound on the lighter stop
mass is then,
\begin{eqnarray}
 m_{\tilde t_1} \lesssim 400~{\rm GeV}\quad (|A_t| \sim m_{\tilde t}),
\end{eqnarray}
\begin{eqnarray}
 m_{\tilde t_1} \lesssim 200~{\rm GeV}\quad (|A_t| \sim 2 m_{\tilde t}),
\end{eqnarray}
for $\Delta^{-1} > 10\%$.

On the other hand, there is no tight naturalness constraint on other
superparticles. The gaugino loop can contribute to $m_{H_u}^2$ or
$m_{\tilde t}$, but the masses can be as large as a few TeV. Other
sfermions can contribute $m_{H_u}^2$ or $m_{\tilde t}$ through the gauge
interactions at two-loop level or through the Yukawa interactions at
one-loop level. Again, masses at a few TeV do not cause a naturalness
problem.

Therefore, the minimal set-up for natural supersymmetry is light stops
and light Higgsinos; we ignore all other superparticles since the
superparticle productions will be anyway dominated by the pair
productions of the lighter stop.
The requirement of the small logarithm implies low energy SUSY breaking
scenarios such as gauge mediation \cite{GMSB1, GMSB2}.  In that case,
there is a nearly massless gravitino to which the lightest MSSM particle
can decay promptly in collider experiments such as the LHC.

This minimal set-up is not particularly motivated by a known UV model of
supersymmetry breaking. It is usually the case that the discussion of
collider signatures of SUSY models requires an assumption of the whole
spectrum and that is why a benchmark model has been needed. However, the
fact that the naturalness consideration predicts a very light colored
particle allows us to postpone the discussion of the detailed structure
of the model or the spectrum. Conversely, one can test the naturalness
principle by looking for a light stop and a light Higgsino.

\section{Natural SUSY at the LHC}\label{sect:NA}

\begin{figure}[t]
  \begin{center}
    \includegraphics[width=8.1cm]{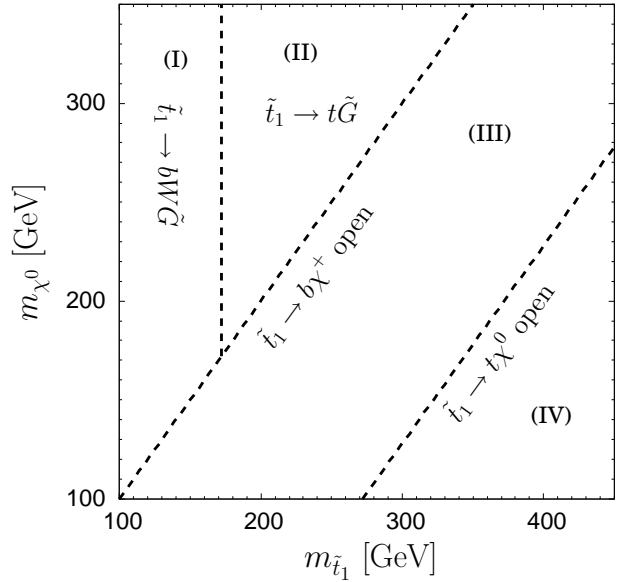}
    \caption{Parameter regions.}
    \label{fig:kinemat}
  \end{center}
\end{figure}

In the study of the stop/Higgsino/gravitino system, there are four
parameter regions which give distinct signatures at the LHC. In
Fig.~\ref{fig:kinemat} we show the four regions (I)--(IV).
If the Higgsino is heavier than the stop and the stop is heavier than
the top quark (region (II)), the stop will decay directly into a top
quark and a gravitino. The final state in this case is $t \bar t$ with
missing momentum. Such events may not be easy to be distinguished from
the $t \bar t$ events. See Refs.~\cite{Meade:2006dw, Alwall:2010jc} for
a study of this event topology. The case with $m_{\tilde t} < m_t$
(region (I)) has been studied in Ref.~\cite{Chou:1999zb}. There the
differences in the distributions of kinematical variables between the
signal and the $t \bar t$ events have been discussed.

In this study, we assume the lighter stop is heavier than the Higgsino,
and the Higgsino decays into the gravitino (regions (III) and (IV)). The
final states are very different from the case with regions (I) and (II).
The main production process is the pair production of the lighter stops,
and then the stops decay into the Higgsinos. The lightest Higgsino
(which we assume to be neutral) in turn decays into $Z$ or $h$ (the
lightest Higgs boson) plus a gravitino.

If the SUSY breaking scale $(\sqrt F)$ is higher than
about 100~TeV, the Higgsino lifetime is long enough to show a displaced
vertex at the level of a sub millimeter.
Such a situation has been studied in Ref.~\cite{Meade:2010ji}, and found
that the LHC can discover the neutralino if $c \tau$ is around
$10^{-1}$--$10^5$ ${\rm mm}$ by looking for displaced $Z$ bosons ($\sqrt
s = 7~{\rm TeV}$ with an integrated luminosity of $1~{\rm fb^{-1}}$).
If the displaced $Z$ bosons are observed, they are a strong indication
of the neutralino/gravitino system.

For $\sqrt F$ smaller than about 100~TeV, which may be reasonable to
assume from the requirement of a low messenger scale, the displaced
vertex of the Higgsino decay will be difficult to observe. In such a
case, the event topology resembles to the case with a stop/Higgsino/Bino
system. We will discuss possible ways to distinguish those two scenarios
later.
In the following analyses, in particular in the mass measurements, we
implicitly make an assumption that the Higgsino decays into a nearly
massless gravitino, not a massive Bino. The assumption should be tested by
other processes or by looking for displaced vertices.

\subsection{Parameters}

In the following we perform a Monte Carlo simulation of the SUSY events
at the LHC. We use the following parameters for the Higgsino/stop
sector:
\begin{eqnarray}
m_{\tilde q_3} = m_{\tilde u_3} = 400~{\rm GeV},
\quad
A_t = -800 ~{\rm GeV},
\end{eqnarray}
\begin{eqnarray}
\mu  = 200 ~{\rm GeV},
\quad
\tan \beta  = 10,
\end{eqnarray}
and we take other superparticle masses to be 1~TeV so that one can
ignore the production of those particles. The mass parameters $m_{\tilde
q_3}$ and $m_{\tilde u_3}$ are the soft SUSY breaking masses for the
left- and right-handed third-generation squarks, respectively.
The relevant mass spectrum and branching fractions calculated from this
parameter set are summarized in Table~\ref{tab:sample_points}. The ISAJET
package~\cite{Paige:2003mg} is used for the calculation of the spectrum
and the branching ratios.
This sample point is in the region (III) in Fig.~\ref{fig:kinemat} where
$\tilde t_1 \to t \chi^0_1$ is closed. The decay mode opens in region
(IV), but since the $\tilde t_1 \to b \chi^+$ decay has a larger
branching ratio in general, the following discussion can also apply 
in region (IV).

\begin{figure}[t]
  \begin{center}
    \includegraphics[origin=c, angle=0,width=7.1cm]{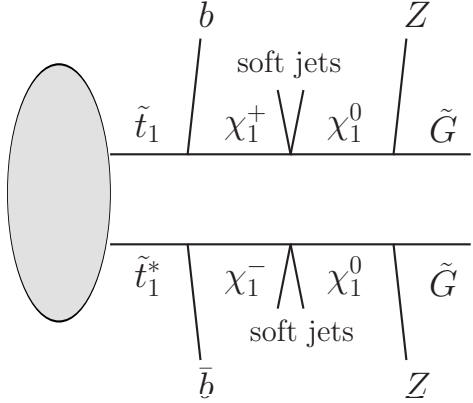}
    \caption{The typical decay chain of the lighter stop.
 }
    \label{fig:stop_decay}
  \end{center}
\end{figure}

The stop masses are 230~GeV and 560~GeV, with which the cross sections
of the squark-pair productions are
\begin{eqnarray}
\sigma_{\tilde{t}_1 \tilde{t}_1} = 18~{\rm pb},
\quad
\sigma_{\tilde{t}_1 \tilde{t}_2}  = 0.10~{\rm pb},
\quad
\sigma_{\tilde{t}_2 \tilde{t}_2}  = 0.11~{\rm pb},
\quad
\sigma_{\tilde{b}_1 \tilde{b}_1}  = 1.1~{\rm pb},
\end{eqnarray}
for $pp$ collisions at $\sqrt s = 14$~TeV.  As one can see, the SUSY
events are dominated by the pair production of the lighter stop. The
typical event is depicted in Fig.~\ref{fig:stop_decay}.
In our analysis, we have produced the hadronized signal events by the
HERWIG event generator~\cite{Corcella:2000bw, Corcella:2002jc}.  For
background processes, we have used ALPGEN~\cite{Mangano:2002ea} and
HERWIG. The detector simulation is based on the AcerDET
package~\cite{RichterWas:2002ch}. We assume 60$\%$ $b$-tagging
efficiency and the mistagging rate to be 10$\%$ (1$\%$) for the charm
quark (other light quarks).

\renewcommand{\arraystretch}{1.3}
\begin{table*}[t]
\center{
  \begin{tabular}{|c||c|ll|}
  \hline
  particle & mass [GeV] & branching ratio & \\
  \hline
  $\chi^0_1$     & 193.8 & Br($\chi^0_1 \to \tilde{G} Z $) = 0.80, &
                           Br($\chi^0_1 \to \tilde{G} h $) = 0.20 \\
  \hline
  $\chi^0_2$     & 202.8 & Br($\chi^0_2 \to \chi^0_1 q \bar{q}$) = 0.40, &
                           Br($\chi^0_2 \to \chi^0_1 \nu \bar{\nu}$) = 0.19, \\
                 &       & Br($\chi^0_2 \to \chi^\pm_1 q \bar{q}$) =
	   0.13, $\cdots$
                         & \\ 
  \hline
  $\chi^+_1$ & 197.3 & Br($\chi^+_1 \to \chi^0_1 q \bar{q}$) = 0.67, &
                       Br($\chi^+_1 \to \chi^0_1 l^+ \nu$) = 0.33 \\
  \hline
  $\tilde{t}_1$  & 230.6 & Br($\tilde{t}_1 \to \chi^+_1 b$)=1.0 & \\
  \hline
  $\tilde{t}_2$  & 559.4 & Br($\tilde{t}_2 \to Z \tilde{t}_1$) = 0.38,& 
                           Br($\tilde{t}_2 \to W^+ \tilde{b}_1$) = 0.20, \\
                 &       & Br($\tilde{t}_2 \to \chi^0_1 t$) = 0.16,& 
                           Br($\tilde{t}_2 \to \chi^+_1 b$) = 0.16, $\cdots$ \\
  \hline
  $\tilde{b}_1$  & 404.1 & Br($\tilde{b}_1 \to W^- \tilde{t}_1$) = 0.73, &
                           Br($\tilde{b}_1 \to \chi^-_1 t$) = 0.24, $\cdots$ \\
  \hline
  $h$            & 119.6 & Br($h \to b \bar{b}$) = 0.82, $\cdots$ &\\
  \hline
  $\tilde{G}$    & $\sim 0$  & &\\
  \hline
  \end{tabular}
}
  \caption{The mass spectrum and the branching ratios used in the
 analysis.}
  \label{tab:sample_points}
\end{table*}
\renewcommand{\arraystretch}{1}

\subsection{Discovery of the light stop}\label{sect:Discovery}

We study discovery potential of the light stop at the early stage of the
LHC experiments. Once the stop pair is produced, they decay into $b
\chi_1^+$ with almost the 100\% branching ratio. The charginos $\chi_1^+$
subsequently undergo the three body decays into $\chi_1^0$ and a quark
or a lepton pair. Finally, $\chi_1^0$ decays mainly into ${\tilde G}
Z$. 
Here $\chi_1^0$ and $\chi_1^+$ degenerate in masses because they are
both Higgsinos. Therefore, the quarks/leptons from the $\chi_1^+$ decays
are rather soft. A typical final state of the stop is then $2b +
2Z + \psl_T$ (Fig.~\ref{fig:stop_decay}).

We take the search strategy for the stop to be to look for $b + Z
(\to l^+ l^-) + \psl_T$.
The main backgrounds are $Z +$ jets and $t \bar{t}$ events. In order to
reduce them, we impose the following selection cuts:
\begin{itemize}
\item at least one $b$-tagged jet with $p_T>30$ GeV,
\item {$M_\mathrm{eff} >$ 350 GeV and $\psl_T >$ 150 GeV},
\item {85 GeV $< m_{l^+l^-} <$ 95 GeV}.
\end{itemize}
Here, $\psl_T$ is a missing transverse momentum and $M_\mathrm{eff}$
is the effective mass~\cite{Hinchliffe:1996iu} which is a scalar sum of
visible and missing transverse momenta.

With an integrated luminosity of 1~${\rm fb}^{-1}$ at $\sqrt{s} = 14$
TeV, the numbers of signal and background events passed through the cuts
are $104$ and $13$, respectively.
At $\sqrt{s} = 7$~TeV, the cross sections for the ${\tilde t_1}{\tilde
t_1}^*$, $t{\overline t}$, and $Z$+jet productions are approximately reduced
by factors of 1/10, 1/6, and 1/3, respectively, compared to those with $\sqrt s
=14$~TeV, {\it i.e.,} approximately 10 events for 2 expected background
event.
Therefore, by looking for this channel, the lighter stop can be discovered
at an early stage of the LHC experiments.

The same final states provide signatures with four leptons. Looking for
this channel will be a non-trivial test of the scenario. The
sensitivities at the LHC is similar to $bZ+\psl_T$.
At the Tevatron, the cross section of the stop pair production is of the
order of 100~fb$^{-1}$ for $m_{\tilde t_1} = 230$~GeV. Considering the
leptonic branching ratio of the $Z$ boson, the four-lepton signature
will be quite challenging to be observed at the Tevatron experiments.

\subsection{Higgsino mass measurement}\label{sect:Higgsino}

Now we turn to the analysis of mass measurements after the
discovery. Although there are two missing gravitinos in each process,
the theoretical input that the gravitino being massless and also the
technique of $M_{T2}$~\cite{Lester:1999tx} help to measure the Higgsino
and stop masses. See Appendix~\ref{app:MT2} for the definition of
$M_{T2}$.

We first discuss determination of the Higgsino mass. The lightest
Higgsino $\chi^0_1$ is mainly produced from the cascade decay of the
stops which are produced in pair. Therefore, in each event, there are a
pair of $\chi^0_1$.
Because of the Higgsino nature, $\chi_1^0$ subsequently decays into $Z
\tilde{G}$ or $h \tilde{G}$.

The $M_{T2}$ variable is suited for this situation as was studied in the
Bino case in Ref.~\cite{Hamaguchi:2008hy}.
We apply the $M_{T2}$ variable for the subsystem $\chi_1^0 \chi_1^0 \to
(Z \tilde{G})(Z \tilde{G})$ $ \to (l^+l^- \tilde{G}) (l'^+l'^-
\tilde{G}) $ in the cascade decays.  
The maximal value of the $M_{T2}$ distribution gives the Higgsino mass
if the true ${\tilde G}$ mass is used in the calculation of $M_{T2}$.
We use the leptonic decay channel of the $Z$ boson in order to reduce
the background from $t \bar t$ productions.

We reconstruct two $Z$ bosons out of four candidate leptons in the final
state. We require the $Z$ candidates to be lepton pairs with the same
flavor and the opposite charges. If all the four leptons have the same
flavor, we take the combination in which the difference of the two
reconstructed $Z$ boson masses is smaller than the other combination.
We impose the following cuts:
\begin{itemize}
\item {four leptons ($p_T > 10$ GeV)},
\item $85 ~{\rm GeV} < m_{l^+l^-} < 95 ~{\rm GeV}$,
\item {$M_\mathrm{eff} >$ 250 GeV and $\psl_T >$ 50 GeV}.
\end{itemize}

\begin{figure}[t]
  \begin{center}
    \includegraphics[origin=c, angle=0,width=7.1cm]{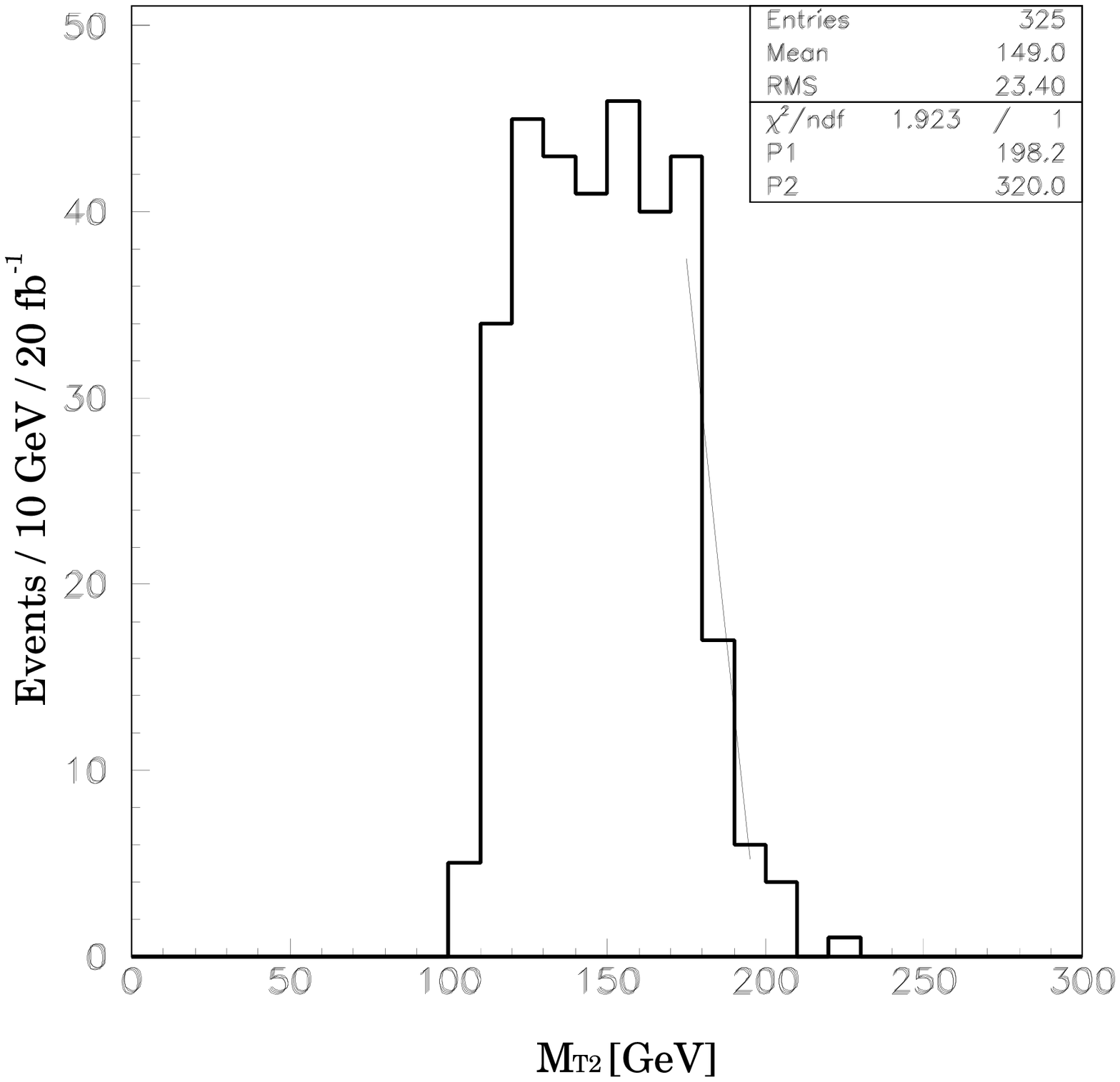}
    \caption{The Higgsino mass measurement. The $M_{T2}$ distribution is
    shown for the $2 Z + \psl_T$ system with an integrated luminosity of
    $\mathcal{L} = 20\, \mathrm{fb}^{-1}$. The endpoint corresponds to
    the Higgsino mass. } \label{fig:Neutralino}
  \end{center}
\end{figure}

In Fig.~\ref{fig:Neutralino}, we show the $M_{T2}$ distribution for the
$2 Z + \psl_T$ system. We used a data set of 20~fb$^{-1}$.
We assume ${\tilde G}$ is massless and thus the maximal value of the
$M_{T2}$ distribution gives the $\chi_1^0$ mass. We can see a clear
endpoint around the input $\chi_1^0$ mass, 193.8 GeV.  As a
demonstration, we fit the endpoint region with a linear function. We
obtained the neutralino mass to be $m_{\chi^0_1} = 198\pm 2 ~{\rm
GeV}$.
The numbers quoted here are sensitive to the choice of the fitting
function and the region to fit. The corresponding error is not included
here.

\subsection{Stop mass measurement}\label{sect:Stop}

One can also measure the stop mass by using the $M_{T2}$ distribution by
including two hard jets (see Fig.~\ref{fig:stop_decay}).
We use a pair of ($bZ$) combinations as visible particles in the
definition of the $M_{T2}$ variable.
Therefore we require at least two and less than five hard jets 
($p_T > 20$~GeV) in addition to the two $Z$ boson candidates (four leptons).
The $Z$ boson candidates are selected in the same way as in the study
of the Higgsino mass measurement.
While attaching a hard jet, we need to decide which $Z$ boson to
combine.
We take a strategy to select a combination which satisfies $m_{j_1
Z_1}+m_{j_2 Z_2} = \min_{i \ne j} (m_{j_i Z_1}+m_{j_j Z_2})$, where
$j_i$ is a label for hard jets.

We only use events in which at least either of $j_1$ or $j_2$ is
$b$-tagged. This selection cut significantly reduces the background from
$2Z +$ jets events. We therefore do not impose the $M_{\rm eff}$ and
the $\psl_T$ cuts in this analysis.

\begin{figure}[t]
  \begin{center}
    \includegraphics[origin=c, angle=0,width=7.1cm]{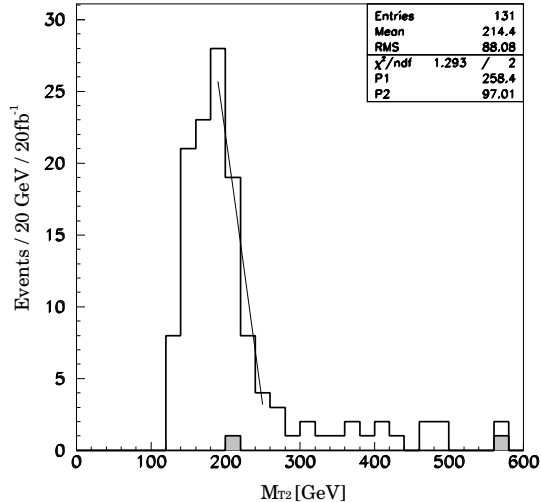}
    \caption{The stop mass measurement. The $M_{T2}$ distribution is
    shown for $2Z + 2j + \psl_T$ system with an integrated luminosity of
    $\mathcal{L} = 20\,\mathrm{fb}^{-1}$. The endpoint corresponds to
    the stop mass.  } \label{fig:Stop}
  \end{center}
\end{figure}

In Fig.~\ref{fig:Stop}, we show the $M_{T2}$ distribution for the $2Z +
2j + \psl_T$ system for an integrated luminosity of 20~fb$^{-1}$.
The hatched histogram is the background from the $2Z + 2j$ events.
Again, by assuming that ${\tilde G}$ is massless, the maximal value
gives the stop mass. We can see a fall off of the histogram around the
input stop mass, 230.6 GeV.  By fitting the endpoint region with a
linear function, we obtain the endpoint to be $m_{\tilde{t}_1} =
258\pm 6~{\rm GeV}$. The error from the choice of the fitting function
and the region to fit is not included here.
%

\subsection{Higgs boson signal}\label{sect:Higgs}

\begin{figure}[t]
  \begin{center}
    \includegraphics[origin=c, angle=0,width=7.1cm]{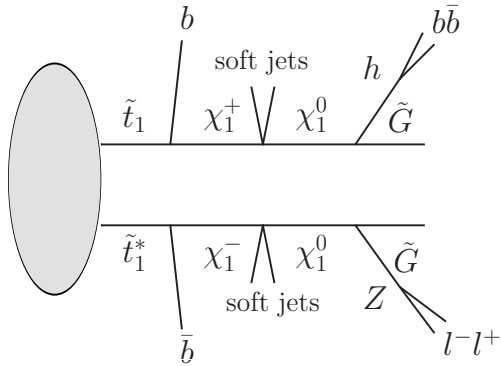}
    \caption{The decay chain for the Higgs boson search.}
    \label{fig:stop_decay_hZ}
  \end{center}
\end{figure}

An interesting signature of the light Higgsino is the production of the
Higgs boson from its decay~\cite{Baer:2000pe, Kitano:2006gv,
Meade:2009qv}.
Here we try to look for an $m_{b\bar b}$ peak from the $h \to b \bar b$
decay (see Fig.~\ref{fig:stop_decay_hZ}).
%
As candidate events we require the following:
\begin{itemize}
\item a lepton pair with the same flavor and opposite charge with $p_T
      > 10$~GeV,
\item 85~GeV $< m_{l^+ l^-} < $ 95~GeV,
\item at least two hard jets with $p_T>50$~GeV,
\item at least three and less than five hard jets with $p_T>30$~GeV,
\item {$M_\mathrm{eff}$ $>$ 250 GeV},
 \item at least one of the two hardest jets is $b$-tagged.
\end{itemize}

There are two difficulties in looking for the Higgs peak. One is the
large $t \bar t$ background which gives a similar final state. The other
is the combinatorial background. There are always additional two
$b$-jets from the stop decays.

The final state of the $t \bar t$ production with leptonic decays of $W$
bosons contains two $b$-jets, two leptons and missing momentum, whereas
the signal event is four $b$-jets, two leptons from a $Z$ boson, and
missing momentum by gravitinos. Although the final state is similar,
there is an effective method to reduce the $t \bar t$ background by
using the $M_{T2}$ variable again.

We treat the two leading $p_T$ jets ($j_1$ and $j_2$) as the Higgs boson
candidates, and the lepton pair ($l_1$ and $l_2$) as the $Z$ boson
candidate.
In order to reduce the $t \bar t$ background, we define the following $M_{T2}$
valuable in this $2l + 2j + \psl_T$ system.
We use $p_{j_1}+p_{l_1}$ and $q_{j_2}+q_{l_2}$ as the two visible momenta 
(which we define the $M_{T2} ((j_1 l_1)(j_2 l_2))$), where the combination
is selected so that it minimizes $m_{j_i l_1}+m_{j_j l_2}$.
If those jets and leptons are from decays of $t \bar t$ pairs, the
$M_{T2}$ variable should have a maximum value at the top quark mass.
Therefore, we impose $M_{T2}((j_1 l_1)(j_2 l_2)) > 180 {\rm GeV}$.

To reduce the combinatorial background, we define another $M_{T2}$
valuable, $M_{T2}((j_1 j_2)(l_1 l_2))$, which has the endpoint at the
neutralino mass if both jets are originated from the decay of the Higgs
boson. We only use events which have the $M_{T2}((j_1 j_2)(l_1 l_2)) <
200 ~{\rm GeV}$.

\begin{figure}[t]
  \begin{center}
    \includegraphics[origin=c, angle=0,width=7.1cm]{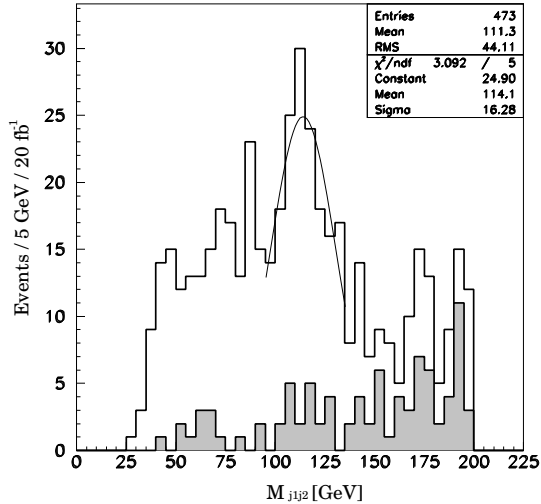}
    \caption{Invariant mass of the $j_1$ and $j_2$ with an 
             integrated luminosity of $\mathcal{L} = 20\,\mathrm{fb}^{-1}$.
 }
    \label{fig:Higgs}
  \end{center}
\end{figure}

In Fig.~\ref{fig:Higgs}, we show the invariant mass distribution of
$j_1$ and $j_2$. The hatched histogram is the background from the $t
\bar t$ events.
As one can see, the background is reduced efficiently by the $M_{T2}$
cuts.
In addition to the peak from the $Z$ boson, we can see a peak of the
Higgs boson. By fitting the peak region by a Gaussian function, 
the Higgs mass is measured to be $m_{h}
=114\pm 16~{\rm GeV}$. The error from the choice of the fitting function
and the region to fit is not included here.

The use of the jet substructure has been proposed in
Ref.~\cite{Kribs:2009yh} to look for boosted Higgs bosons from the
neutralino decays. Although the Higgs bosons in this present study are
not very boosted, the method may be useful to make the peak sharper.

\section{Neutralino vs. Gravitino}

Here we comment on the question of how we confirm that the invisible
particle is indeed the nearly massless gravitino rather than another
neutralino such as the Bino.
This question will be important after the discovery of the light stop
through the $bZ$ + $\psl_T$ channel.
The clearest signal for the gravitino hypothesis will be the displaced
vertex of the Higgsino decay. In the following we consider a case when
SUSY scale is too low to observe the displaced $Z$ bosons.

In the case of the prompt decay, in principle, the mass of the missing
particle can be measured by combining another independent quantity with
the present $M_{T2}$ analyses.
For example, the endpoint of the $m_{jZ}$ invariant mass distribution in
the stop decay (see Fig.~\ref{fig:stop_decay}) can be such a quantity.
The endpoint value is expressed in terms of masses:
\begin{eqnarray}
m_{jZ}^{\rm max} 
&=& m_{\tilde{t}_1}
\sqrt{\left( 1 - \frac{ m_{\chi^0_1}^2 }{ m_{\tilde{t}_1}^2 } \right)
      \left( 1 - \frac{ m_{\tilde{G}}^2 }{ m_{\chi^0_1}^2 }   \right)
      }.
    \label{fig:invM_endpoint}
\end{eqnarray}
In the $m_{\tilde{G}} \to 0$ GeV limit, $m_{jZ}^{\rm max} = (
m_{\tilde{t}_1}^2 - m_{\chi^0_1}^2)^{1/2}$.
This combination is, however, similar to the one we can extract from the
$M_{T2}$ analysis, and thus the constraints we obtain will not be enough
to claim the almost massless gravitino.
Within the study of the stop pair production, it seems that it remains
as a problem to distinguish two scenarios which are drastically
different; one with high scale supersymmetry breaking and the other with
low scale supersymmetry breaking.

A better way is to look for a Drell-Yan process for the Bino-Higgsino
pair production. For a light Bino and a Higgsino there will be
significant cross section at the LHC and at the ILC. The final state is
$Z$ with a missing momentum without hard jets. The lack of such events
will be an indication of the gravitino scenario.
Another clear distinction is possible when photon channels are available.
If the neutralino (mostly Higgsino) has a small Bino component, we
expect $2b+Z+\gamma+\psl_T$ and $2b+2\gamma+\psl_T$. Since the
Higgsino/Bino system can give the photon signatures only through a loop
diagram, a large number of the photon signal will be a clear indication
of low scale supersymmetry breaking.

\section{Summary}\label{sect:Summary}

The LHC experiments are running to look for new physics. 
If we take the naturalness seriously in supersymmetric models, there
should be a rather light scalar top quark and a Higgsino. Also the
requirement of the low messenger scale from naturalness suggests that
there is a nearly massless gravitino. The rest of the particles are
useless for naturalizing the electroweak symmetry breaking, and
therefore can be as heavy as a few TeV. In such a case, the LHC
signatures are quite different from the conventional studies.

The approach taken in this paper can be generalized to other models.
Taming the one-loop corrections to the Higgs potential always needs a
new particle $T^\prime$ which is the stop in supersymmetric case. The
presence of $T^\prime$ within the LHC reach would be a robust
prediction.
We assumed in the study that $T^\prime$ decays into $b + Z + \psl_T$. If
that is the signal, it is clear enough to be distinguished from the
standard model processes at the early stage of the LHC experiments.

We have limited our discussion to the MSSM, but the requirements of the
light stop and the light Higgsinos are quite general in supersymmetric
models whereas the Higgs boson can be much heavier than 120~GeV in
extended models. In that case, the Higgsino decay into the Higgs boson
is further suppressed or forbidden.
We have constructed a simplified model representing a class of models
which solve the naturalness problem.
The signal of natural supersymmetry is a clean $2b + 2Z + \psl_T$. This
scenario will be discovered/excluded quite soon at the LHC.

\section*{Acknowledgements}
This work was supported in part by the Grant-in-Aid for the Global COE
Program, ``Weaving Science Web beyond Particle-matter Hierarchy,'' from
the Ministry of Education, Culture, Sports, Science and Technology of
Japan (M.A.), KRF-2008-313-C00162 (H.K.), NRF with CQUEST 2005-0049409
(H.K.), and the Grant-in-Aid for Scientific Research 21840006 of JSPS
(R.K.). RK and HK thank the organizers of the meeting at SLAC,
``Workshop on Topologies for Early LHC Searches,'' September 22-25,
2010, where authors had numbers of useful conversations with
participants.

\appendix
 \section{Definition of $M_{T2}$}\label{app:MT2}

 In this section, we define a useful valuable,
 $M_{T2}$~\cite{Lester:1999tx}.  We consider a system in which two
 particles are pair-produced (which we call ``$A$") and each of them
 decays subsequently into an invisible particle $X$ and visible ones. In
 the system, $AA \to \{ {\rm visible}(p) X(k) \} + \{ {\rm
 visible}(p^{\prime}) X(k^{\prime}) \} $ , the $M_{T2}$ variable is
 defined by
 \begin{eqnarray}
 M_{T2}
 &=& \min_{ {\bf k}_T + {\bf k}_T^{\prime} = \psl_T } 
     \left[ \max \left\{ M_T ({\bf p}_T, {\bf k}_T), 
                         M_T ({\bf p}^{\prime}_T, {\bf k}^{\prime}_T) \right\} \right],
     \label{fig:MT2_1}
 \end{eqnarray}
 where $\psl_T$ is the missing momentum and $p$($p^{\prime}$) is a sum
of momenta of visible particles, $p = \sum_i p_i$ ($p^{\prime} = \sum_i
p^{\prime}_i$).  The transverse mass, $M_T$, is defined by
 \begin{eqnarray}
  M_T^2 ({\bf p}_T, {\bf k}_T)
 &=& m_{\rm visible}^2 + m_X^2 + 
     2 \left( E^{\rm visible}_T E^X_T 
             - {\bf p}_T \cdot {\bf k}_T
      \right).
     \label{fig:MT2_2}
 \end{eqnarray}
If we postulate the true value of $m_X$ in the above formula, we can
 obtain a parent particle mass $m_A$ as the endpoint of $M_{T2}$.

\end{document}